\documentclass[review,12pt]{elsarticle}

\usepackage{amssymb}
\usepackage{graphicx}
\usepackage{times}
\usepackage{newtxmath}
\usepackage{xcolor}
\usepackage{multirow}
\usepackage{array}
\usepackage{lineno}
\usepackage{algorithm}
\usepackage[noend]{algpseudocode}
\usepackage{amsmath,amsfonts}
\usepackage{adjustbox}

\journal{Computational Materials Science}

\begin{document}

\begin{frontmatter}

\title{\textbf{Neural Network Accelerated Process Design of Polycrystalline Microstructures}}

 \author[1]{Junrong Lin}
 
 \author[2]{Mahmudul Hasan}

\author[2]{P{\i}nar Acar}

\author[3]{Jose Blanchet}

\author[1]{Vahid Tarokh}

 \address[1]{Department of Electrical and Computer Engineering,
Duke University, Durham, NC 27708 USA}

 \address[2]{Department of Mechanical Engineering, Virginia Tech, Blacksburg, VA 24060 USA}

\address[3]{Department of Management Science and Engineering, Stanford University, CA 94305 USA}

\begin{abstract}
Computational experiments are exploited in finding a well-designed processing path to optimize material structures for desired properties. This requires understanding the interplay between the processing-(micro)structure-property linkages using a multi-scale approach that connects the macro-scale (process parameters) to meso (homogenized properties) and micro (crystallographic texture) scales. Due to the nature of the problem's multi-scale modeling setup, possible processing path choices could grow exponentially as the decision tree becomes deeper, and the traditional simulators' speed reaches a critical computational threshold. To lessen the computational burden for predicting microstructural evolution under given loading conditions, we develop a neural network (NN)-based method with physics-infused constraints. The NN aims to learn the evolution of microstructures under each elementary process. Our method is effective and robust in finding optimal processing paths. In this study, our NN-based method is applied to maximize the homogenized stiffness of a Copper microstructure, and it is found to be 686 times faster while achieving 0.053\% error in the resulting homogenized stiffness compared to the traditional finite element simulator on a 10-process experiment.

\end{abstract}

\end{frontmatter}

\section{Introduction}
The research on investigating process-structure-property relationships has become more prevalent with the introduction of the Integrated Computational Materials Engineering (ICME) paradigm~\cite{allison2006integrated}, which aims to solve complex and multi-scale material design problems. As a result of the recent advancements following the introduction of ICME, several aspects of computational materials science and process engineering have significantly improved. For instance, new methodologies are developed for ICME to lower the costs and risks associated with the processing of new materials~\cite{cowles2015update,venkatesh2018icme}. Computational experiments are exploited to eliminate the traditional trial-error approach in bridging material features and properties. The investigation of the process-(micro)structure-property linkages requires a multi-scale approach that connects the macro-scale (process parameters) to meso (homogenized properties) and micro (crystallographic texture) scales. Physics-based multi-scale modeling of materials provides an opportunity to achieve the optimum design of materials with enhanced properties. However, their high computational cost prevents these multi-scale approaches from being widely adopted and used by industry in real material design efforts~\cite{harrington2022application}. Nevertheless, the process-structure problem is studied less than the structure-property problem as the physics behind the process-structure problem is generally more complicated. Machine learning tools have demonstrated promise to address this gap by building low-cost process-structure-property surrogate models to replace computationally expensive physics-based models~\cite{brough2017microstructure}.

Control of polycrystalline microstructures is important in material design, processing, and quality control since the orientation-dependent material properties (e.g., stiffness) could change as the underlying microstructures evolve during a deformation process. Therefore, identification of the optimal processing route to produce a material with desired texture and properties plays an important role in materials research. Traditional experimental approaches used to explore the optimum processing routes for a given texture are usually based on trial and error and, thus, can be tedious and expensive. Hence, computational methods are developed to replace these experiments to accelerate the design of microstructures with desired textures. Fast and accurate prediction of texture evolution during processing can significantly contribute to linking the current material design and manufacturing efforts for polycrystalline materials.

Current research in this field is focused on process-structure~\cite{sarkar2018implementing, yabansu2017extraction, tapia2017bayesian, cecen2018material, hassinger2016toward, Acar16} or structure-property~\cite{acar2016utilization,acharjee2003proper,ganapathysubramanian2004design,kalidindi2004microstructure,kalidindi2000prediction,wargo2012selection,fullwood2010microstructure} relationships separately. For example, Sarkar et al.~\cite{sarkar2018implementing} developed a surrogate model for $ZrO_{2}$-toughened $Al_{2}O_{2}$ ceramics to predict the sinter density and grain size from sintering heat treatments process parameters. In another study, Tapia et al.~\cite{tapia2017bayesian} established a Gaussian process regression-based surrogate model for the heat treatment process of NiTi shape-memory alloys where the input parameters are heat treatment temperature and its duration, as well as initial nickel composition, and the output is the final nickel composition after heat treatment. Next, Acar and Sundararaghavan~\cite{acar2016utilization} developed a linear solver-based multi-scale approach using reduced-order modeling to design target microstructures that optimize homogenized material properties. In order to learn the reduced basis functions that can adequately represent the crystallographic texture of polycrystalline materials,  Achargee et al.~\cite{acharjee2003proper} and Ganapathysubramanian et al.~\cite{ganapathysubramanian2004design} used the proper orthogonal decomposition (POD) and method of snapshots in Rodrigues orientation space~\cite{morawiec1996rodrigues} to create a continuum sensitivity-based optimization technique to compute the material properties that are sensitive to the microstructure. Similarly, Kalidindi et al.~\cite{kalidindi2004microstructure} designed the microstructure for a thin plate with a circular hole at the center to maximize the uniaxial load-carrying capacity of the plate without plastic deformation. There are few other studies available in the literature that explore the process-structure and structure-property relationships for additively manufactured materials~\cite{yan2018modeling,hashemi2022computational,popova2017process}. Very recently, Dornheim et al.~\cite{dornheim2022deep} developed a model-free deep reinforcement learning algorithm to optimize the processing paths (up to 100 combinations) for a targeted microstructure. Their algorithm does not require prior samples rather it can connect with the processing simulations during optimization. They also extended the method to solve multi-objective optimization problems. In another study, Honarmandi et al.~\cite{honarmandi2022accelerated} proposed a novel framework based on batch Bayesian optimization to solve the inverse problem to find the material processing specifications using microstructure data. Using both low-fidelity and high-fidelity phase field models, they developed a Gaussian process regression-based surrogate model to replace the computationally expensive process models and integrated it into inverse design optimization.

\begin{figure}[b!]
\centering
\includegraphics[trim={6.5cm 2cm 0cm 1cm},clip,width=1.5\textwidth]{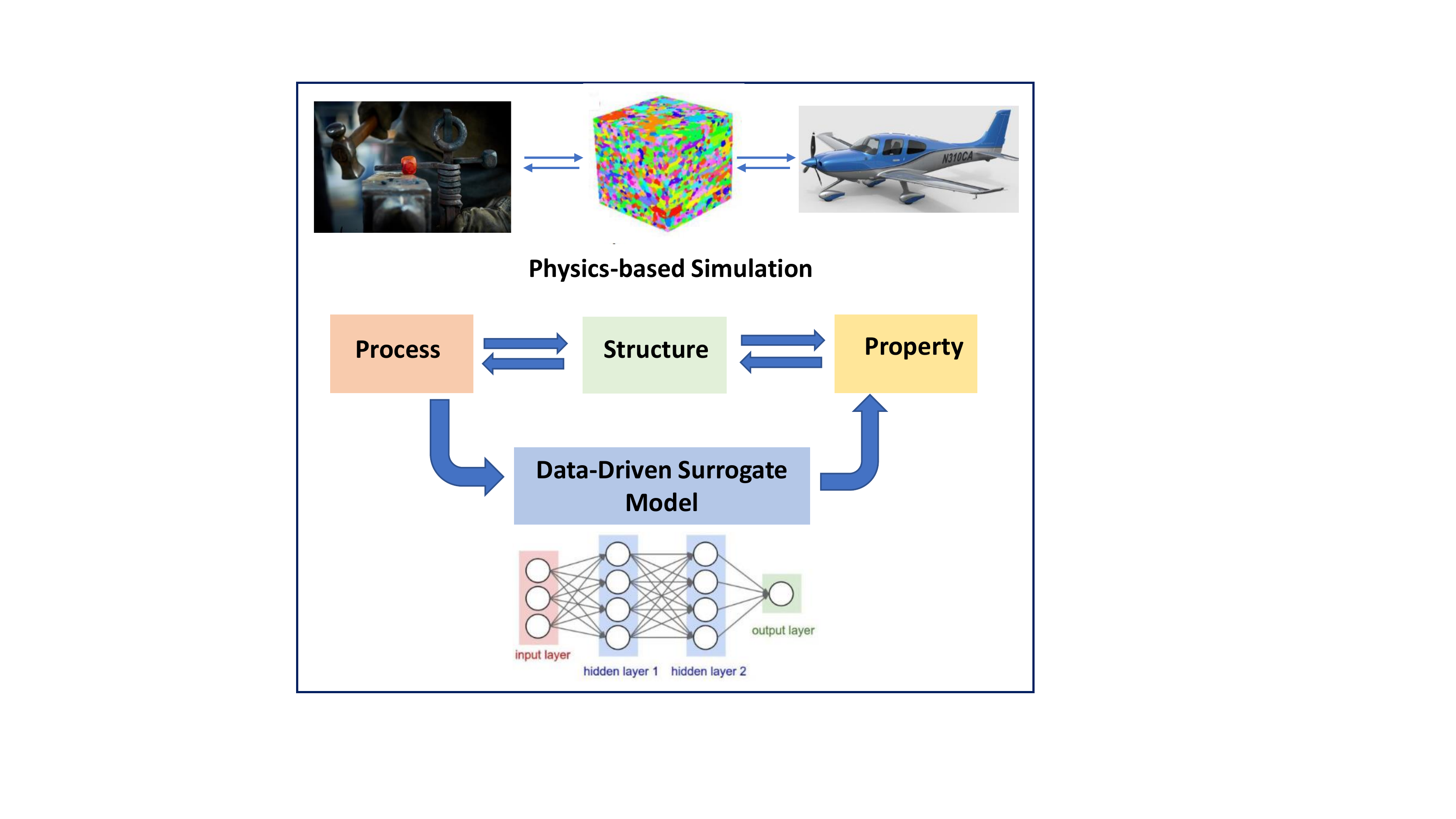}
\caption{Schematic of the contribution of this study. Data-driven surrogate model is developed to replace the physics-based simulator on the process-structure-property problem.}
\label{fig:overview}
\end{figure}

Recent studies~\cite{Acar16,zabaras05} model the texture evolution of polycrystalline materials using a probabilistic representation called orientation distribution function (ODF), which describes the volume density of crystals of different orientations in a microstructure. Predicting the changes of the microstructural texture after applying a particular deformation process (e.g. applying shear force along a particular direction for 1 second) involves solving the ODF conservation equation, which is a differential equation~\cite{zabaras05, kumar00}. The traditional physics-based solution is developed using the finite element method to solve the conservation equation numerically~\cite{kumar00}. The change in the ODFs during processing also controls the homogenized (meso-scale) mechanical properties of the microstructures. Significant challenges arise when we carry out a large-scale search task to find out microstructures with desired orientation-dependent properties. Under this context, we expect to optimize some homogenized mechanical properties by sequentially applying different deformation processes. The optimal processing path could be found by searching algorithms coupled with a process simulator of microstructure evolution. However, in practice, even a simple task (e.g. given an arbitrary initial texture and 5 different deformation processes and their combinations, finding out the exact optimal path of applying 10 processes sequentially) could be time-consuming to solve using traditional physics-based simulators due to both simulation time requirements and exponentially growing number of possible processing paths. Concurrent multi-process modeling can further increase the computational complexity of the material design. Simulation speed becomes the bottleneck when confronting these tasks. Therefore, this study develops a supervised learning approach together with a local search algorithm to explore and bridge the process-microstructure-property relationships of polycrystalline materials .

The neural network has already been shown to be a revolutionary function approximator in this era of abundant data which is already adopted in the literature~\cite{hasan2022machine,mann2022development,han2011prediction,abbod2002physically,fang2009approach} to study process-structure-property linkages in order to build low-cost surrogate models. Inspired by physics-informed neural networks (PINN)~\cite{raissi19}, we will develop a surrogate neural network trained on a small dataset to replace the physics-based simulator in this process design task. The overall contribution of this study is summarized in Fig.~\ref{fig:overview}. The high-speed inference feature of a pre-trained neural network could accelerate the algorithm and therefore enable large-scale searching. Compared to the finite element (FE) simulator, this method is faster and has a little trade-off in prediction accuracy. The organization of this article is as follows: Section 2 and Section 3 discuss the mathematical formulation for modeling of deformation processing and polycrystalline microstructures, respectively. Section 4 and Section 5 describe the physics-based and data-driven modeling of the process-microstructure relationship, respectively. The performance of the neural network-based surrogate model of process design for improved mechanical performance is reported in Section 6. Finally, Section 7 provides a summary of the paper and a discussion on potential future work.  

\section{Mathematical Modeling of Deformation Process with ODF Approach}\label{sec:phy}

Multiple crystals with various crystallographic orientations make up a polycrystalline material, and these orientations determine the microstructural texture, which is mathematically described by the orientation distribution function (ODF). During a deformation process, the texture of polycrystalline microstructure changes under applied loads. Deformation process modeling with the ODF approach is computationally efficient compared to the expensive finite element solver. The ODF, denoted by $A(\mathbf{r},t)$, indicates the volume density of the crystals in the orientation space, $\mathbf{r}$ in a certain time $t$. ODFs are discretized over the Rodrigues orientation space using finite element techniques. The reduced region, called the fundamental region $\Omega$, is induced based on the crystallographic symmetry of the polycrystal system from the initial Rodrigues orientation space. The Rodrigues angle-axis parameterization method is used to depict the various crystal orientations. In contrast to the Euler angles representation~\cite{Bunge,Wenk}, this method uses axis-angle representations to express crystal orientations. The Rodrigues parameterization is described by scaling of the axis of rotation, $\mathbf{n}$, as $\mathbf{r}=\mathbf{n}tan(\frac{\theta}{2})$, where $\theta$ is an angle of crystal rotation. For further information on Rodrigues parameterization of microstructural solution spaces, interested readers are referred to Refs.~\cite{Acar16, kumar00}.\\  The ODF, $A(\mathbf{r},t)$, could be used to compute homogenized elastic stiffness $ <C>$ through its volume integration over the fundamental region, $\Omega$:

\begin{equation} \label{eq:1}
     <C> = \int_{\Omega} C(\mathbf{r}) A(\mathbf{r}, t)\,dV
\end{equation}

where \emph{C}(\textbf{r}) includes the single-crystal material properties required to compute the homogenized elastic stiffness values given by $<C>$. By controlling the ODF values, desired homogenized (meso-scale) properties can be obtained as described in Eq. (\ref{eq:1}). However, the ODF ($A(\mathbf{r}, t)\geq 0$) must satisfy the volume normalization constraint which is expressed as follows:

\begin{equation}
     \int_{\Omega} \ A(\mathbf{r},t) \,dV=1
     \label{eq.6}
 \end{equation}

The homogenized (meso-scale, volume-averaged) properties of the microstructures are obtained using the given expression in Eq. (\ref{eq:1}). Here, the integration for the homogenized properties is performed over the fundamental region by considering the lattice rotation, $\textbf{R}$. Given the Rodrigues orientation parameter, $\mathbf{r}$, the rotation, $\textbf{R}$, can be obtained with the following expression:

\begin{equation}
\textbf{R} = \frac{1}{1+\mathbf{r} \cdot \mathbf{r}}(I(1-\mathbf{r} \cdot \mathbf{r})+2(\mathbf{r}\otimes \mathbf{r} + I \times \mathbf{r}))
\end{equation}

\begin{figure}[htb]
\centering
\includegraphics[trim={6cm 5cm 2cm 2.5cm},clip,width=1.3\textwidth]{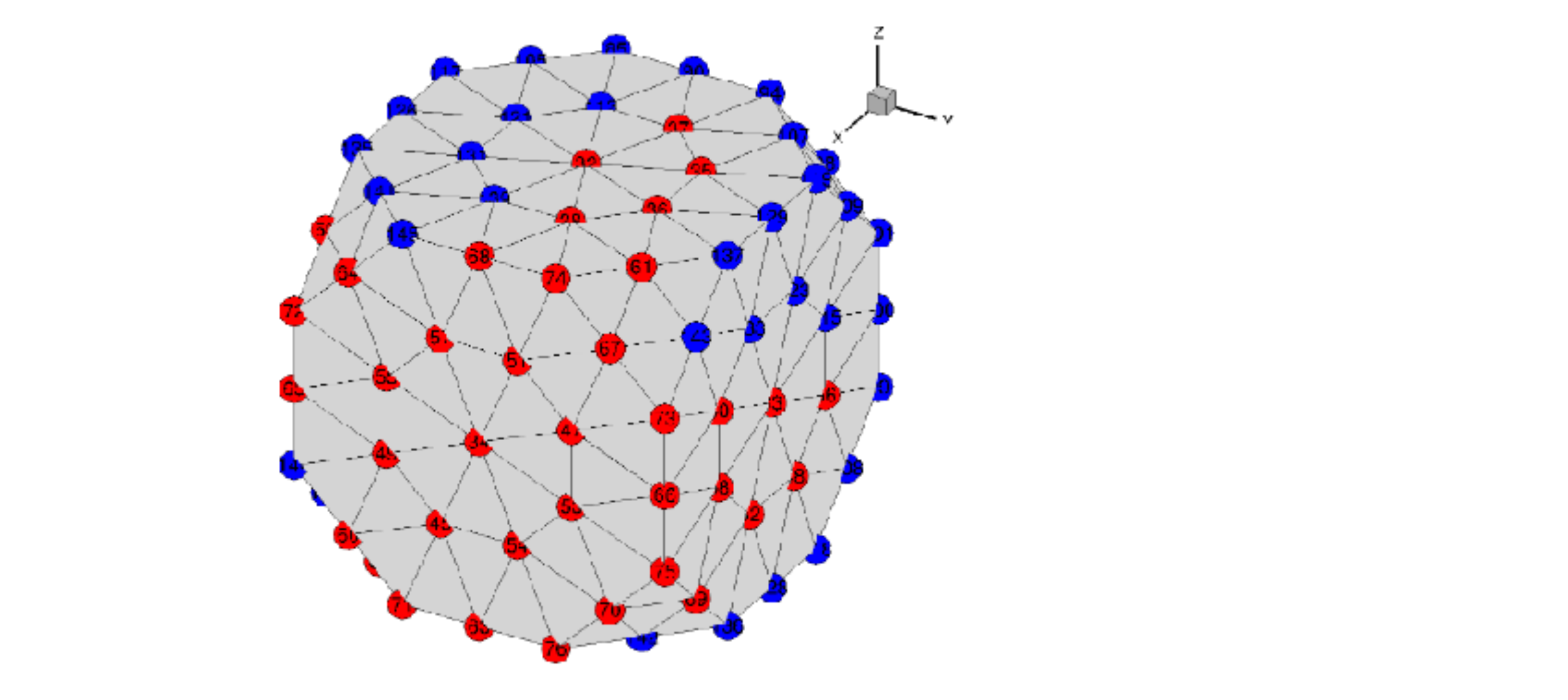}
\caption{Finite element discretization of the orientation space for face-centered cubic (FCC) microstructures. The red-colored nodal points show the independent ODF values while the blue-colored nodes indicate the dependent ODFs as a result of the crystallographic symmetries.}
\label{fig:mesh}
\end{figure}

The finite element discretization of the microstructural orientation space is exhibited in Fig. \ref{fig:mesh}. Here, each independent nodal point of the finite element mesh represents a unique ODF value for the associated crystallographic orientation. For $N$ independent nodes with $N_{elem}$ in the finite element discretization ($N_{int}$ integration points per element), Eq. (\ref{eq:1}) can be approximated as follows at a given time, $t$:
 
 \begin{equation}
\begin{aligned}
   <C>&=\int_{\Omega}C(\mathbf{r}) A(\mathbf{r}, t) \,dV\\
    &=\sum_{n=1}^{\ N_{elem}}\sum_{m=1}^{\ N_{int}}   C(\mathbf{r_m}) A(r_m)\omega_m|J_n|\frac{1}{(1+r_m.r_m)^2}  
    \label{eq.9}
\end{aligned}
\end{equation}

where \(A(r_m)\) is the ODF value at the \(m^{th}\) integration point with global coordinate \(r_m\) (orientation vector) of the \(n^{th}\) element. \(|J_n|\) is the Jacobian determinant of the \(n^{th}\) element and \(\omega_m\) is the integration weight of the \(m^{th}\) integration point. 

Considering the crystallographic symmetry, the homogenized property of Eq. (\ref{eq.9}) can alternatively be approximated in the linear form as \(\textcolor{red}{<C>}=P^{T}\textbf{A}\), where $P$ is the property matrix that is a product of the single-crystal material properties and finite element discretization of the orientation space, and $\textbf{A}$ is the column vector of the ODF values for the independent nodes (this work uses 76 independent nodes to model the FCC microstructure) of the finite element mesh (see Fig. \ref{fig:mesh}). Moreover, the ODF should satisfy the normalization constraint (see Eq. \ref{eq.6}). The normalization constraint is mathematically equivalent to the fact that the sum of the probabilities for having all possible crystallographic orientations in a microstructure must be one.

 During the deformation process, ODFs change due to the reorientation of the grains. It evolves from the initial ODFs (t=0) to the final deformed ODFs (t=t). The evolution of the ODF values is governed by the ODF conservation equation, which satisfies the volume normalization constraint of Eq.~\ref{eq.6}. Equation (\ref{eq:2}) shows the Eulerian rate form of the conservation equation in the crystallographic orientation space with a gradient operator given by \cite{zabaras05}:

\begin{equation} \label{eq:2}
    \frac{\partial A(\mathbf{r}, t)}{\partial t} + \nabla A(\mathbf{r},t) \cdot v(\mathbf{r}, t) + A(\mathbf{r}, t)\nabla\cdot v(\mathbf{r}, t)=0
\end{equation}

where $v(\mathbf{r}, t)$ is the reorientation velocity.\\ 
The texture evolution can be calculated by the microstructure constitutive model in terms of a velocity gradient (\textbf{L}) definition (see Eq.~\ref{eq.10} below), which is linked to $v(\mathbf{r}, t)$ by the Taylor macro-micro linking hypothesis. The Taylor hypothesis assumes that the crystal velocity gradient is equal to the macro velocity gradient~\cite{zabaras05}.
To compute the reorientation velocity, a rate-independent constitutive model is adopted. Resulting $A(\mathbf{r}, t)$ (current texture) which evolves from $A(\mathbf{r}, 0)$ (initial texture) is solved by utilizing the constitutive model and finite element representation of the Rodrigues orientation space.

Each deformation process (e.g., tension/compression and shear) yields a particular ODF as output after applying the load for a certain time. The macro velocity gradient, \textbf{L}, for a particular process is provided as input in the crystal plasticity solver to investigate the ODF evolution during that process. While designing a process sequence to obtain the desired texture, the macro velocity gradient describing the processing route (type and sequence of the process and strain rate) is solved as a design variable. The velocity gradient of a crystal with the orientation, $\mathbf{r}$, can be expressed as

\begin{equation}
   \textbf{L}=S+\textbf{R}\sum_{\alpha}\dot{\gamma^{\alpha}}\Bar{\textbf{T}}^{\alpha}\textbf{R}^T,
   \label{eq.10}
\end{equation}

\noindent where $S$ represents the lattice spin, $\textbf{R}$ indicates the lattice rotation, $\dot{\gamma^{\alpha}}$ and $\Bar{\textbf{T}}^{\alpha}$ indicate the shearing rate and Schmid tensor for the slip system $\alpha$, respectively. Here, the crystal velocity gradient is defined in terms of the time rate change of the deformation gradient. Equation (6) demonstrates that the velocity gradient is decomposed into two components. The first component, lattice spin $S$, is related to the elastic deformation gradient by assuming that the deformation gradient ($\textbf{F}$) is decomposed into elastic ($\textbf{F}^e$) and plastic ($\textbf{F}^p$) parts such that: $\textbf{F} = \textbf{F}^e \textbf{F}^p$. In particular, the lattice spin is equal to $\dot{\textbf{R}^e}\textbf{R}^e$, where $\textbf{R}^e$ is evaluated through the polar decomposition of the elastic deformation gradient: $\textbf{F}^e = \textbf{R}^e \textbf{U}^e$, where $\textbf{U}^e$ is the unitary matrix of the polar decomposition \cite{acar2018computational}. The second component in Eq. (6) shows the rotated plastic velocity gradient ($\textbf{R} \textbf{L}^p \textbf{R}^T$). Here, $\textbf{L}^p$ denotes the plastic velocity gradient, which is related to the combined shearing of the slip systems given by $\sum_{\alpha}\dot{\gamma^{\alpha}}\Bar{\textbf{T}}^{\alpha}$.

The macro velocity gradient expression of Eq.~(\ref{eq.10}) for different deformation processes can also be written in the following matrix form, given in Eq.~(\ref{eq.11}). The detailed derivation from Eq.~(\ref{eq.10}) to Eq.~(\ref{eq.11}) is skipped here for brevity and it can be found in Ref.~\cite{zabaras05}.

\begin{multline}
    \textbf{L}=\alpha_{1}\begin{bmatrix} 1 & 0 & 0\\ 0 & -0.5 & 0\\ 0 & 0 & -0.5
    \end{bmatrix}+\alpha_{2}\begin{bmatrix} 0 & 0 & 0\\ 0 & 1 & 0\\ 0 & 0 & -1
    \end{bmatrix}+\alpha_{3}\begin{bmatrix} 0 & 1 & 0\\ 1 & 0 & 0\\ 0 & 0 & 0
    \end{bmatrix}\\+\alpha_{4}\begin{bmatrix} 0 & 0 & 1\\ 0 & 0 & 0\\ 1 & 0 & 0
    \end{bmatrix}+\alpha_{5}\begin{bmatrix} 0 & 0 & 0\\ 0 & 0 & 1\\ 0 & 1 & 0
    \end{bmatrix}
    \label{eq.11}
\end{multline}

Each matrix in  Eq.~(\ref{eq.11}) defines a deformation process, e.g. tension/compression ($\alpha_{1}$), plane strain compression ($\alpha_{2}$), and shear modes ($\alpha_{3}, \alpha_{4}, \alpha_{5} $).

We define a processing path of \emph{n} steps with a fixed deformation duration time $\Delta$\emph{t} in an autoregressive pattern below. $\phi_{F_i}(A)$ stands for a deformation process with load  $F_i$ (e.g., tension/compression and shear modes defined through the velocity gradient, $\textbf{L}$, in the finite element simulations) for a duration $\Delta$\emph{t} on a polycrystalline microstructure described by the ODF, $A(r, i\Delta t)$, at time \emph{i$\Delta$t}:
\begin{equation} \label{eq:3}
\begin{split}
    A(\mathbf{r}, \Delta t) &= \phi_{F_0}(A(\mathbf{r}, 0)) \\
    A(\mathbf{r}, 2\Delta t) &= \phi_{F_1}(A(\mathbf{r}, \Delta t)) \\
    &\vdots \\
    A(\mathbf{r}, n\Delta t) &= \phi_{F_{n-1}}(A(\mathbf{r}, (n-1)\Delta t))
\end{split}
\end{equation}
Here $F_i\in \mathcal{F}$ is picked from a pre-specified set $\mathcal{F} \coloneqq\{f_1, \dots, f_k\}$ (e.g. a set of shear force $\{xy, xz, yz\}$) of deformations we can apply in each step. An optimal processing path $P^*$ for an initial ODF $A(\mathbf{r},0)$ is a sequence of deformation processes that could achieve the best ODF set $A^*(\mathbf{r},n\Delta t)$ at time \emph{n$\Delta$t}. $A(\mathbf{r}, n\Delta t)$ is evaluated by the desired orientation-dependent properties which is an integral of \emph{A} and single-crystal material property matrix (weight function) $C(\mathbf{r})$ over the fundamental region $\Omega$ (as shown in Eq. (1)). 
\begin{equation}
\begin{split}
    \underset{A}{\max}\ F = \sum_{i}^6 \omega_{ii} <C_{ii}> + \  \sum_{i<j}^6 \omega_{ij} <C_{ij}> \\
    \text{subject to} \int_{\Omega} A(\mathbf{r}, i\Delta t)dV = 1& \\
    A(\mathbf{r}, i\Delta t) \geq 0&
\end{split}
\end{equation}

where the composite objective function ($F$) includes the summation of the constants of the $6 \times 6$ anisotropic homogenized elastic stiffness matrix ($<C_{ii}>$, $<C_{ij}>$) that are computed using the formulation given in Eq. (1). In this problem, the diagonal entries of the homogenized stiffness matrix ($<C_{ii}>$) are assumed to be more important, and thus multiplied by a weight factor of $\omega_{ii} = 1$, while the cross-diagonal terms ($<C_{ij}>$) are multiplied by a weight factor of $\omega_{ij} = \frac{1}{2}$. Note that the symmetric terms ($<C_{ij}>$ and $<C_{ji}>$) are only counted once (through the $i<j$ condition in the second term of the objective function) in the summation formula. The underlying idea behind this composite objective function is to improve the homogenized elastic stiffness of the microstructure in different directions rather than improving the stiffness along a particular direction.

To address this task, we need 1) an efficient deformation simulator $\phi_F(A)$ and 2) a path-searching algorithm. For the algorithm, various works have been done on such local search problems. These algorithms work by making a sequence of decisions locally to optimize the objective function, with well-known methods including simulated annealing. In this work, we focus on a novel approach to building up $\phi_F(A)$ by a neural network. As a function approximator, a neural network could learn the behavior of different deformations accurately from data and predict fastly to accelerate process design. It is chosen to address the challenge arising from the numerically intractable nature of the processing design problem and the speed bottleneck of a traditional simulator.




\section{End-to-end deformation prediction with Neural Networks}

We aim to develop an efficient neural network to replace the FE predictor in computing deformation results. In this data-driven method, we no longer focus on solving the conservative equation presented in Section \ref{sec:phy} but exploit some physical constraints to build up a surrogate neural network. The neural network aims to approximate the deformation process $\phi_F$:
\begin{equation}
    A(\mathbf{r}, \Delta t) = \phi_F(A(\mathbf{r},0)),\ \ \mathbf{r}\in{\Omega}
\end{equation}
where $A(\mathbf{r}, 0)$ denotes the ODFs before process and $A(\mathbf{r}, \Delta t)$ is the ODFs after applying a deformation \emph{F} of duration $\Delta$\emph{t}.

We define ${NN}_F(A;\theta)$ to be the surrogate neural network model with $\theta$ parameters. In this work, we employ multilayer perceptron (MLP) to approximate $\phi_F(A)$. For an MLP with $L$ hidden layers, we have:
\begin{equation}
\begin{split}
    a^{(0)} &= x\\
    z^{(i+1)} &= M^{(i)}a^{(i)}+b^{(i)}\\
    a^{(i+1)} &= \varphi^{(i+1)}(a^{(i)}) = \xi^{(i+1)}(z^{(i+1)}) \\
    NN_F &= \varphi^{(L)} \circ\cdots\circ \varphi^{(1)} 
\end{split}
\end{equation}
where \emph{x} is the model input, $\xi^{i+1}$ denotes the activation function of layer \emph{i}. Considering that the ODF stands for the probability density of the orientation space, the ODF non-negativity and volume normalization constraints must be satisfied. A ReLU followed by a normalization layer which divides each channel with a material-specific weighted sum between volume fraction constant and output from the previous layer are applied to keep the network outputs physically feasible.

The model parameters can be learned by minimizing the error between exact ODFs and network predictions. One common choice is using squared error to match them. Here we introduce the specific objective value weights to assign different importance to different elastic stiffness constants. Considering that the integral of the homogenized properties in Eq. (\ref{eq:1}) over a discrete fundamental is well approximated the weighted sum (Eq. (\ref{eq.9})), the weighted mean squared error (WMSE) should be a more appropriate choice:

\begin{equation}
    WMSE = \frac{1}{B}\frac{\sum_{i=1}^{n}w_i\Vert y_i-y^*_i \Vert^2_2}{\sum_{i=1}^{n}w_i}
\end{equation}
where y and y* stand for true and predicted ODF values, respectively, $w_i$ are the orientation weights to calculate the desired objective material property, and \emph{B} is batch size.


 In this section, we evaluate our method on a stiffness optimization task. We find a minimal difference in predicted ODFs and objective property value, compared with the traditional FE method. In addition, the neural network predicts hundreds of times faster than traditional methods which enables large-scale searching.

\subsection{Model Setup} \label{result:1}
We consider a stiffness optimization task of finding out the best texture evolution path by applying 31 different possible deformation processing modes (tension, compression, and shear along xy, xz, and yz axes, respectively, and their combinations, except the case when none of them are selected). The fundamental region is discrete with 76 possible outcomes (the FCC microstructure is modeled with 76 independent ODF values). Only one of the deformation processes can be applied in one step. Each step is 0.1 seconds in duration. An evolution path consists of 10 sub-processes executed in order. The elastic stiffness is selected as the objective mechanical property to be maximized while the material of interest is Copper (Cu). The following single crystal properties are taken for Copper (Cu): $C_{11}=C_{22}=C_{33}=$ 168 GPa, $C_{12}=C_{21}=C_{13}=C_{23}=C_{31}=C_{32}=$ 121.4 GPa, and $C_{44}=C_{55}=C_{66}=$ 75.4 GPa~\cite{zabaras05}. The objective function is defined as the sum of the homogenized elastic stiffness constants ($C$) using a composite function definition with weights of 1 for diagonal entries ($C_{ii}$) and weights of 0.5 for off-diagonal entries ($C_{ij}$, $i \neq j$) of the elastic stiffness matrix. The symmetric off-diagonal ($C_{ij}$, $i \neq j$) entries are only considered once. The synthetic dataset of size 5000 is uniformly initialized and normalized to keep the ODFs feasible according to the unit volume normalization constraint (Eq. (2)). The deformation mode results of each initial ODF are generated by the FE solver. In the following experiment, data points used for training and testing are selected randomly. The dataset is also randomly divided by an 80\%/20\% train-test ratio.

\subsection{Searching Algorithm and network structure}
Due to the complex nature of the combinational optimization problem, calculating all the $31^{10}$ results is impractical even with fast NN. Instead, we employ a heuristic algorithm that helps to find out the optimal process path by sampling from a multinominal distribution constructed from the predicted value at each stage.

\begin{algorithm}[H]\small
\caption{Exponential Weights Algorithm}\label{alg:cap}
\begin{algorithmic}
\Require $\text{RESTART} = N$, $\text{STEPS} = M$, BASE $ = \beta$
\For{$i = 1 \text{ to } N$}
\State $A_0 = initial\ ODF$
\For{$j = 1 \text{ to } M$}
\State Simulate 31 possible outcomes by neural network
\State $A_1[1:31] \gets NN_F(A_0)$
\State $\text{obj\_value\_diff} \gets W(A_1[1:31]-A_0)$
\If{$\text{obj\_value\_diff}\preceq0$}
\State \textbf{break}
\Else
\State $weights \gets \text{SoftMax}(\text{obj\_value\_diff};\beta)$
\State $k \gets multinominal(weights)$
\State $PATH[i][j] \gets k$
\State $A_0 \gets A_1[k]$
\EndIf
\EndFor
\EndFor
\State Return the best solution found
\end{algorithmic}
\end{algorithm}

The neural network has 1 hidden layer of 760 neurons with hyperbolic tangent activation functions except the last ReLU. ReLU and normalization are applied to fit the physical constraint in the last layer. The network is trained with a mini-batch of size 128. ADAM with warm restarts~\cite{LoshchilovH19} is used for network training. When searching, restart 1000 times with weight base $\beta=5$ for each initial texture.

\subsection{Network performance and searching samples} \label{result:3}

 \begin{figure}[ht!]
\centering
\includegraphics[trim={3cm 3cm 0cm 2cm},clip,width=1.1\textwidth]{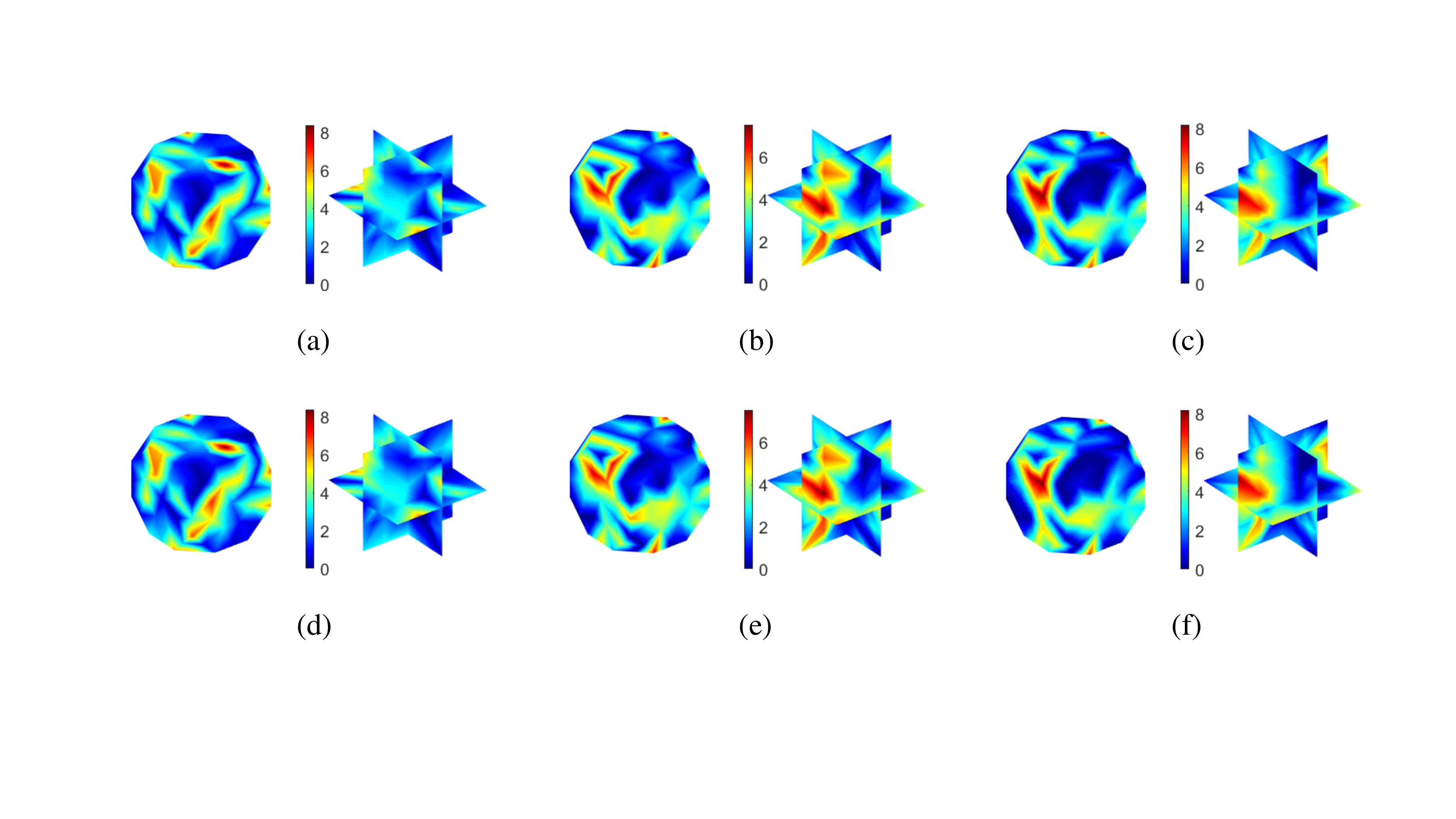}
\caption{The comparison between the predicted textures by the neural network model (a, b and c) and finite element crystal plasticity model (d, e and f) at time steps t=0.3 sec (a and d), t=0.8 sec (b and e), and t=1 sec (c and f).}
\label{fig:comp}
\end{figure}

\begin{figure}
    \centering
    \includegraphics[scale=0.5]{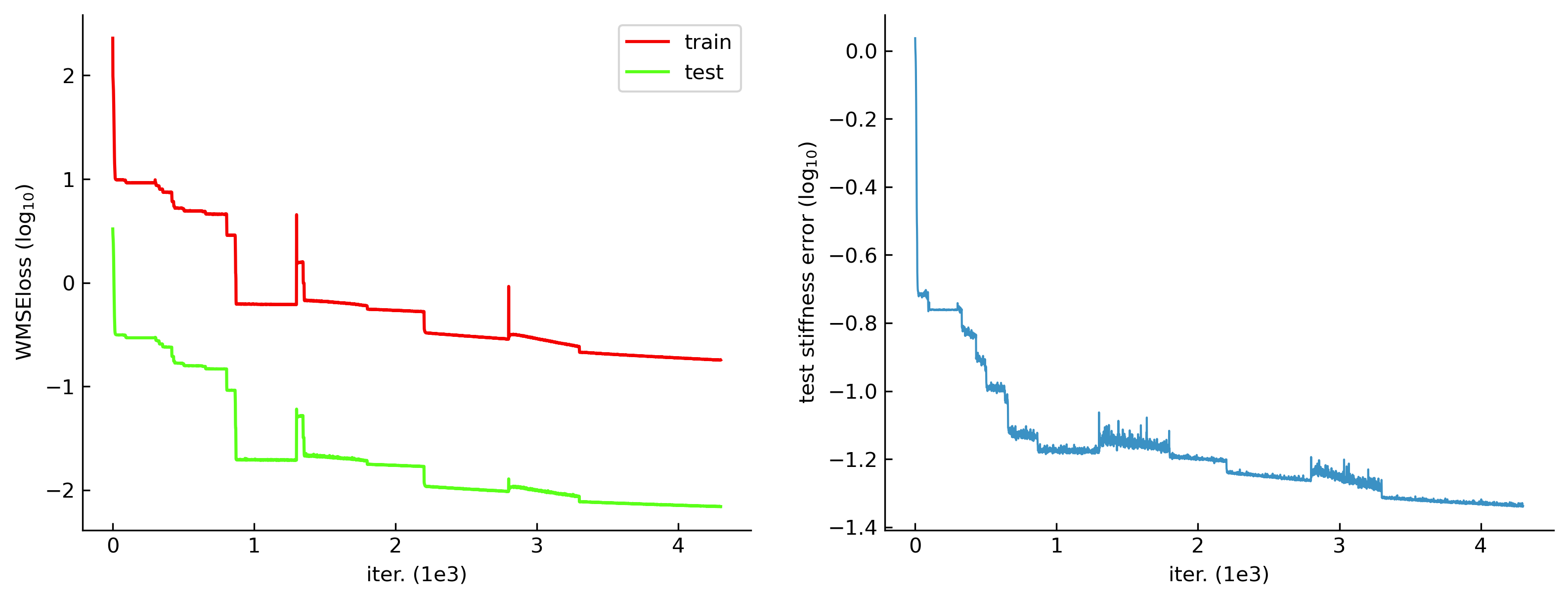}
    \caption{Training error and testing error (left) on the synthetic dataset. Average stiffness error amount 31 modes (right).}
    \label{fig:loss}
\end{figure}

Figure~\ref{fig:comp} depicts the comparison between the textures obtained from the neural network prediction and FE simulator at different time steps of processing. The loading conditions are as follows: for Fig.~\ref{fig:comp} (a) and (d) shear force in the YZ plane, for Fig.~\ref{fig:comp} (b) and (e) a combined force of tension, compression, and shear in XZ and YZ planes, and for Fig.~\ref{fig:comp} (c) and (f) a combined force of tension, compression, and shear in the YZ plane. The performance of the neural network is also evident from Fig. \ref{fig:loss} and Table \ref{table:L2_stiff}, which presents the relative average errors of independent ODFs over the fundamental region for different deformation modes.

\begin{table}[H]
\vspace{12pt}
\adjustbox{max width=\textwidth}{%
\centering
 \begin{tabular}{||*{8}{c}||}
 \hline
 10000 & 01000 & 00100 & 00010 & 00001 & 11000 & 10100 & 10010\\ [0.1ex] 
 \hline
 0.034\% / 0.041 & 0.036\% / 0.048 & 0.041\% / 0.048 & 0.038\% / 0.046 & 0.039\% / 0.048 & 0.044\% / 0.052  & 0.035\% / 0.044 & 0.038\% / 0.041 \\
 \hline\hline
 10001 & 01100 & 01010 & 01001 & 00110 & 00101 & 00011 & 11100\\
 \hline
 0.046\% / 0.053 & 0.033\% /0.044 & 0.044\% /0.051 & 0.041\% / 0.045 & 0.045\% / 0.051 & 0.040\% / 0.045 & 0.043\% / 0.048 & 0.0314\% / 0.043 \\
 \hline\hline
 11010 & 11001 & 10110 & 10101 & 10011 & 01110 & 01101 & 01011\\
 \hline
 0.043\% /0.045 & 0.0476\% / 0.049 & 0.0347\% / 0.044 & 0.0323\% / 0.043 & 0.037\% / 0.047 & 0.033\% / 0.043 & 0.033\% / 0.042 & 0.041\% /0.044 \\
 \hline\hline
 00111 & 11110 & 11101 & 11011 & 10111 & 01111 & 11111 & \\
 \hline
 0.035\% / 0.049 & 0.030\% / 0.039 & 0.031\% / 0.041 & 0.045\% / 0.051 & 0.034\% / 0.046 & 0.033\% / 0.046 & 0.032\% / 0.044& \\
 \hline
 \end{tabular}} 
 \caption{The average relative L2 error/stiffness error of neural network predictions compared with FE simulator on each deformation mode. We use binary strings of length 5 to represent different surrogate networks. Each digit represents one of the deformation processing modes (tension, compression, and xy, xz, yz shear). 1 if it is applied, otherwise 0 (e.g., 10010 represents the network that learns the deformation process when tension and xz shear are simultaneously applied).} \label{table:L2_stiff}
\end{table}

The computation time for predicting all 31 deformation modes' results from an arbitrary set of initial ODF values in the neural network is 0.2213s while the traditional finite element simulator takes 152s on average when running in a 31 process-based parallelism. The neural network gives predictions 686.85 times faster. The predictions of the neural network were generated on NVIDIA(R) Tesla P100-16G with single-core Intel(R) Xeon(R)@2.3Ghz and simulator run on AMD(R) EPYC C2D-highcpu-32@2.45 Ghz. The average difference in the elastic stiffness prediction values of the physics-based model and neural network after 10 processes is 0.4663 GPa.

\begin{figure}[b!]
\centering
\includegraphics[trim={0cm 0cm 0cm 0cm},clip,width=1.1\textwidth]{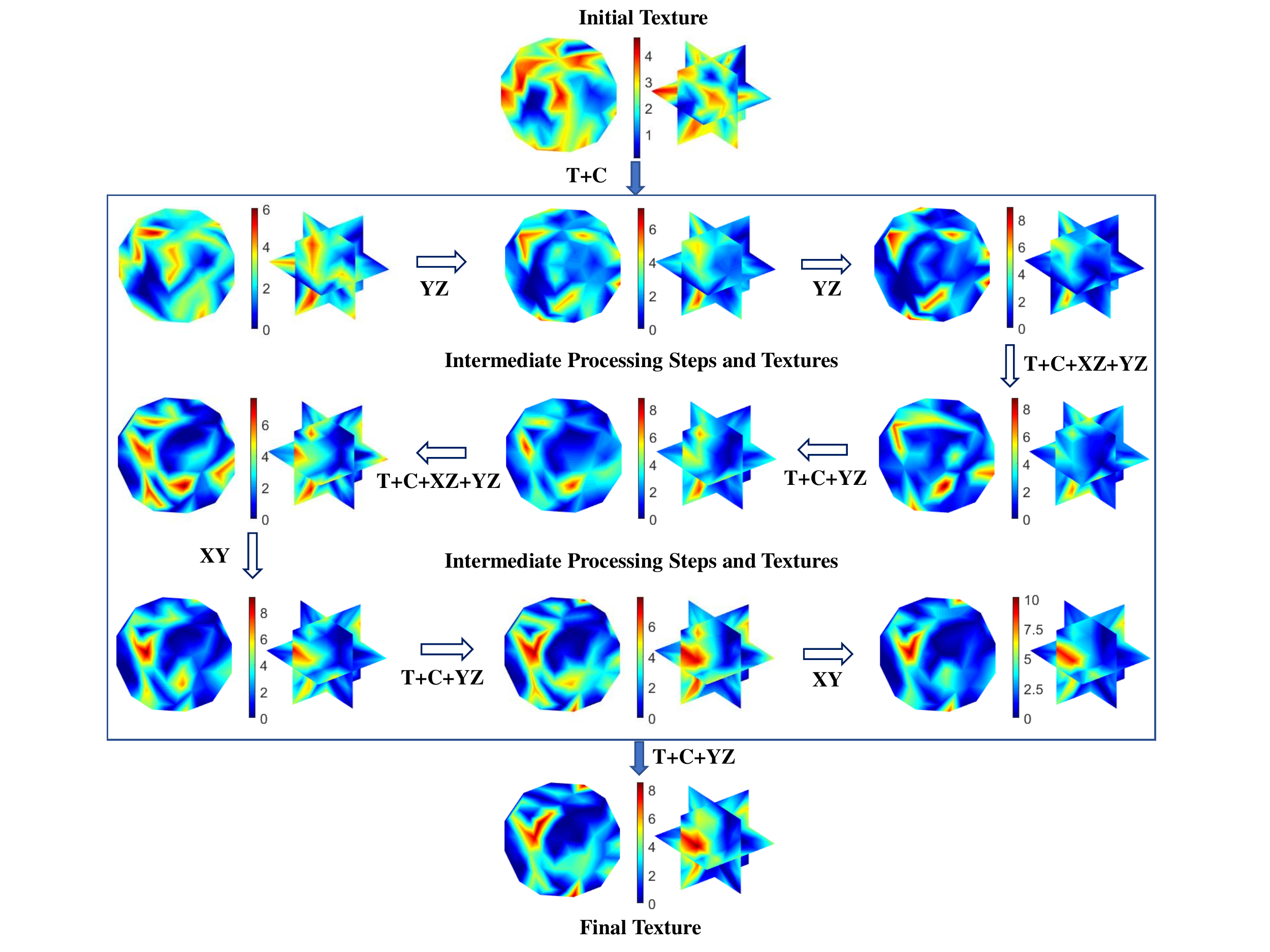}
\caption{Texture evolution prediction through the optimum processing path by the neural network surrogate model. The figure shows the different steps of deformation processing from an initial texture to a final optimum texture which maximizes an objective function defined for the homogenized elastic stiffness constants of a Copper microstructure. T and C stand for tension and compression, respectively, and XY, XZ \& YZ represent the corresponding shear processes.}
\label{opt:NN}
\end{figure}

\begin{figure}[b!]
\centering
\includegraphics[trim={0cm 0cm 0cm 0cm},clip,width=1.1\textwidth]{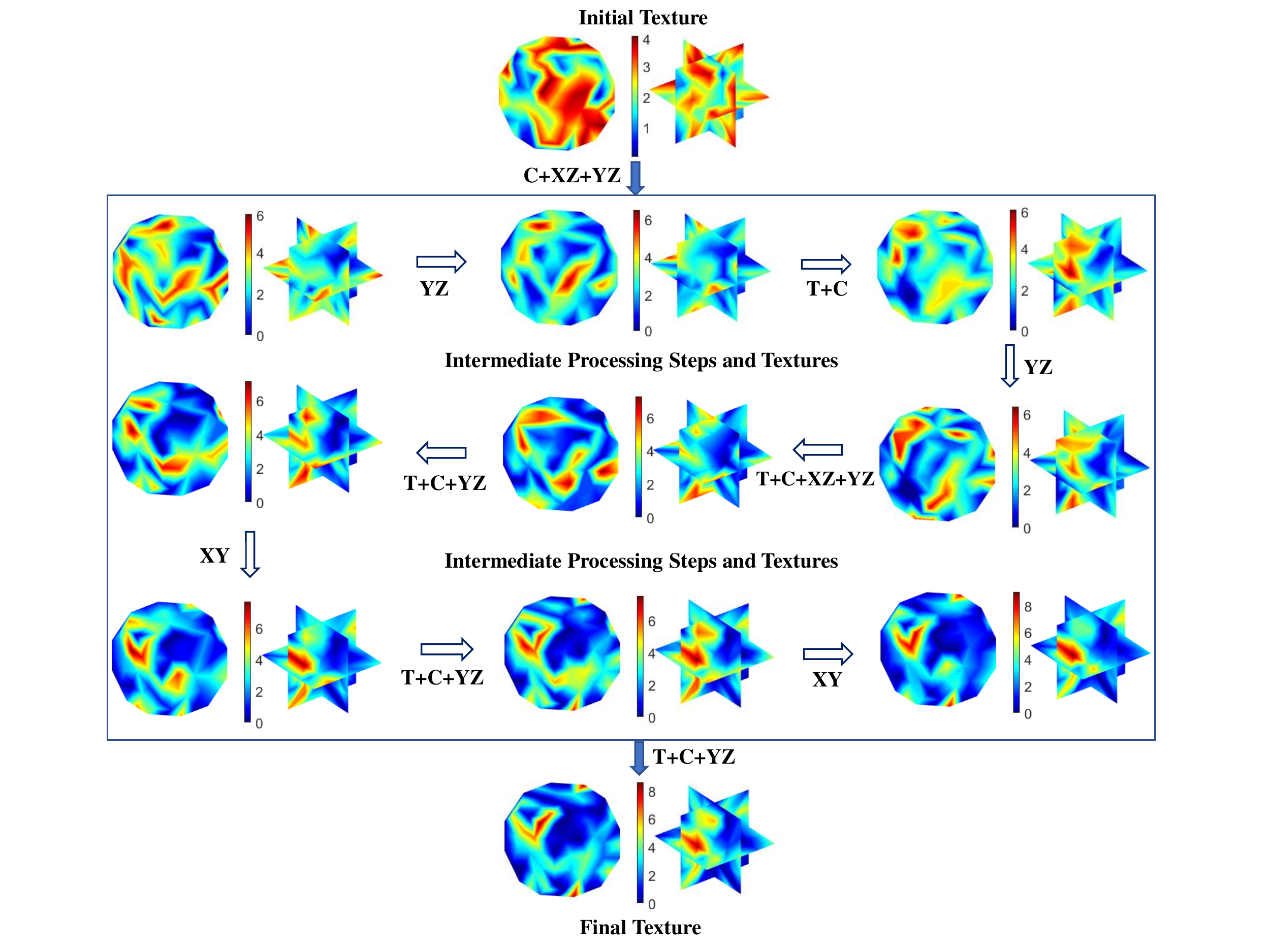}
\caption{Texture evolution through the optimum processing path by the physics-based simulator. The figure shows the different steps of deformation processing from an initial texture to a final optimum texture which maximizes an objective function defined for the homogenized elastic stiffness constants of a Copper microstructure. T and C stand for tension and compression, respectively, and XY, XZ \& YZ represent the corresponding shear processes.}
\label{opt:PB}
\end{figure}

The neural network-based surrogate model replaced the physics-based finite element model to find the optimum processing paths. It is already mentioned that an optimization problem is developed with the goal of maximizing the sum of the elastic stiffness constants of Copper. The solution of the optimization problem provides a maximum objective function value with the corresponding texture. It also suggests an optimum processing route to obtain the optimum texture. We determine the processing route for five different initial textures. Though the optimum solutions vary with the initial textures, almost equal objective function values are obtained for all cases. The average of the sum of the maximum stiffness constants of Copper is found to be 885.2 GPa from the neural network surrogate model whereas this value is 884.9 GPa for the FE-based process model. 

\begin{figure}[b!]
\centering
\includegraphics[trim={5cm 4cm 3cm 3cm},clip,width=1.2\textwidth]{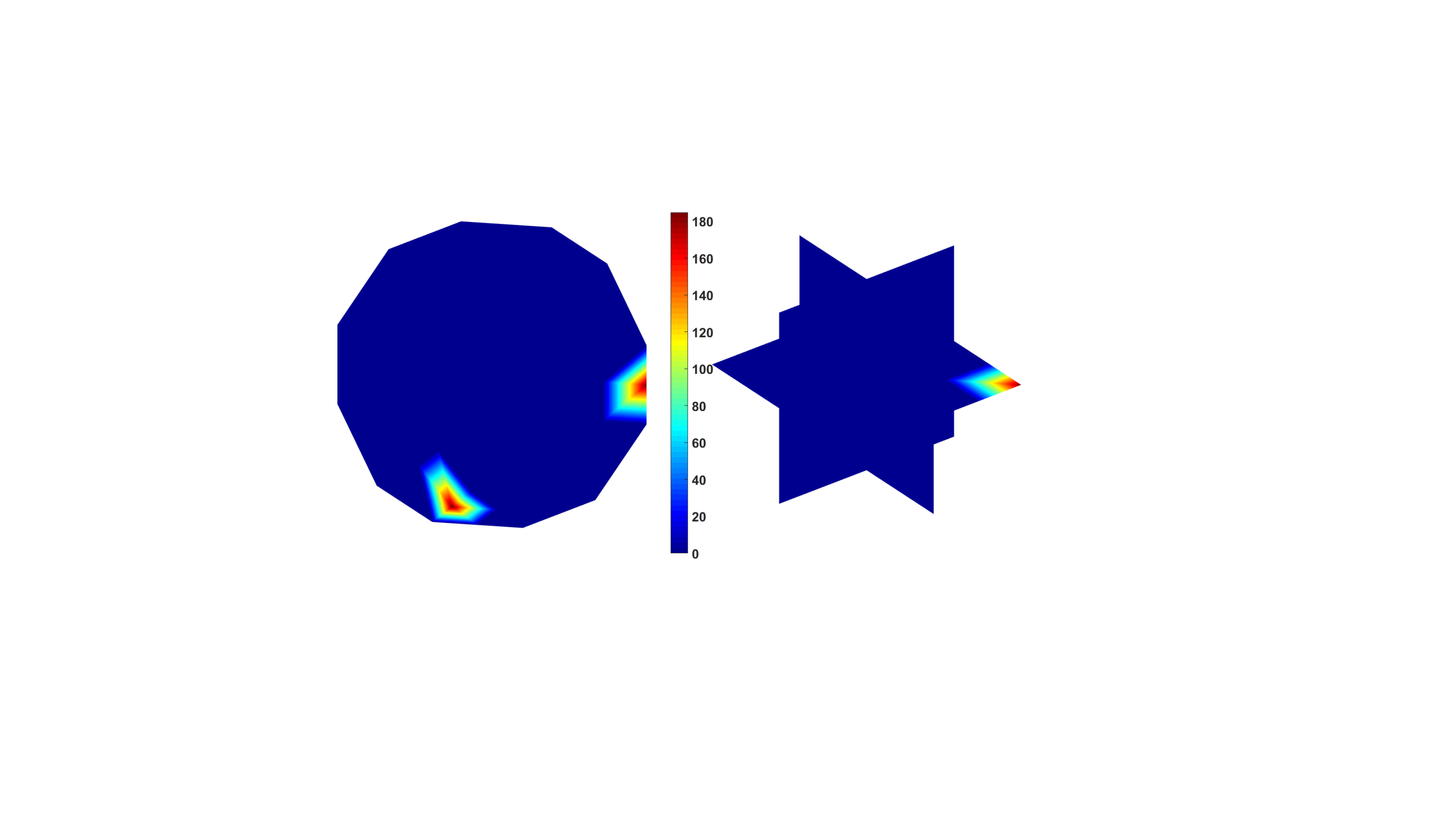}
\caption{A single crystal optimum texture is obtained using linear programming to maximize the homogenized elastic stiffness constants without considering processing.}
\label{opt:single}
\end{figure}

The neural network predicted the optimum processing path with texture evolution after each step to obtain the optimum texture at the end starting from a certain initial texture, as shown in Fig.~\ref{opt:NN}. Similarly, the optimum process route and texture evolution obtained from the physics-based model for another initial texture is displayed in Fig.~\ref{opt:PB}. Here, T and C stand for tension and compression, respectively, and XY, XZ \& YZ represent the corresponding shear processes. The strain rate is constant for all steps (i.e., 1 $s^{-1}$). The processing is allowed to have only one loading condition or combined loading conditions as reported in Fig.~\ref{opt:NN} and Fig.~\ref{opt:PB}. Both models are found to provide polycrystalline textures as optimum solutions. 

If we solve the same optimization problem using linear programming without considering the processing (only structure-property problem), a single crystal texture is found as an optimum ODF solution (see Fig.~\ref{opt:single}). In this case, the maximum objective function value is 896.2 GPa, which indicates the theoretical maximum value of the objective function. On the other hand, the neural network surrogate model, which accounts for processing (process-structure-property problem), provides an optimum texture with 885.2 GPa for the objective function value (which is a higher value compared to the randomly oriented texture providing an objective function value of 878.4997 GPa). However, the theoretical solution of the structure-property problem results in a single crystal solution while the solutions are polycrystalline textures for the process-structure-property problems using both the neural network and physics-based model. Even though single crystal textures provide the theoretically possible maximum value for elastic stiffness constants, their manufacturing is difficult. On the other hand, the polycrystalline textures of the process-structure-property problem can easily be manufactured with the presented simple deformation processing modes. Therefore, considering the effects of the processing is significant for manufacturing and bridging materials design and manufacturing~\cite{hasan2022data}.

This work can be a valuable addition to the literature to investigate the process-structure-property linkages of polycrystalline microstructures. The neural network surrogate model can be a substitute for the physics-based simulator as it is computationally faster and accurate. Moreover, the processing-produced optimum textures predicted by the surrogate model can still improve the mechanical properties (e.g. elastic stiffness) near to the maximum theoretical values.

\section{Conclusions}

In this work, we developed a surrogate neural network model to accelerate the process design of polycrystalline microstructures. The traditional physics-based FE simulator was found inefficient to address the large-scale searching task. With the representation potential of the neural network, a pre-trained network could predict the ODF after deformation significantly fast and accurately. An example design problem was solved to find the optimum processing route maximizing an objective function defined for the homogenized elastic stiffness constants of Copper. The results demonstrate a good match between the predictions of the physics-based simulator and neural network surrogate model. Studies on accumulative error analysis of neural network are reserved in the future. Future work may also involve the integration of a similar data-driven modeling strategy for the concurrent multi-scale modeling of metallic components.

\section{Acknowledgements}

Material in this paper is based upon work supported in part by the Air Force Office of Scientific Research under award number FA9550-20-1-0397. MH and PA also acknowledge the support from the Air Force Office of Scientific Research Young Investigator Program under grant FA9550-21-1-0120 and from the National Science Foundation under award number 2053840.


\section{Data availability}
The raw/processed data required to reproduce these findings cannot
be shared at this time as the data also forms part of an ongoing study.
The data will be made available upon request.

\bibliography{ref.bib}
\bibliographystyle{elsarticle-num}

\end{document}